\documentclass[layout=onecolumn, journal=jpcafh]{achemso}

\setkeys{acs}{chaptertitle = true}

\usepackage{geometry}
\usepackage{graphicx,subfigure,mciteplus}
\usepackage{amssymb, amsmath}
\usepackage[flushleft]{threeparttable}
\usepackage{epstopdf, booktabs}
\usepackage{multirow,bigdelim}
\usepackage{ragged2e}
\makeatletter
\g@addto@macro\TPT@defaults{\footnotesize}
\makeatother

\usepackage{graphicx}
\usepackage{dcolumn}
\usepackage{bm}
\usepackage{longtable}
\usepackage{multirow}

\newcommand{\AS}{[$N_a,r_a$]}
\newcommand{\six}{[6, 6]}   
\newcommand{\ten}{[10, 10]}
\newcommand{\fourteen}{[14, 14]}
\newcommand{\eighteen}{[18, 18]}
\newcommand{\twentytwo}{[22, 22]}
\newcommand{\twentysix}{[26, 26]}
\newcommand{\thirty}{[30, 30]}
\newcommand{\FeMoco}{[Mo:7Fe:9S:C]:(R)-homocitrate}
\newcommand{\femoco}{MoFe$_7$S$_9$C}

\newcommand{\moietyone}{MoFe$_3$S$_7$}
\newcommand{\moietytwo}{Fe$_4$S$_7$}
\newcommand{\rdm}{$^2D$}

\def\bra#1{\bigl<#1\bigr|}
\def\ket#1{\bigl|#1\bigr>}

\def\ahatd#1{\hat{a}^\dagger_{#1}}
\def\ahat#1{\hat{a}_{#1}}

\author{Jason M. Montgomery}
\affiliation{Department of Chemistry, Biochemistry, and Physics, Florida Southern College, Lakeland, FL 33801}
\author{David A. Mazziotti}
\email{damazz@uchicago.edu}
\affiliation{Department of Chemistry and The James Franck Institute,
University of Chicago, Chicago, IL 60637}

\title[V2RDM Calculations of FeMoco]{Strong Electron Correlation in Nitrogenase Cofactor, FeMoco}

\begin{document}

%
%
%
%
%

\begin{abstract}
FeMoco, \femoco, has been shown to be the active catalytic site for the reduction of nitrogen to ammonia in the nitrogenase protein. An understanding of its electronic structure including strong electron correlation is key to designing mimic catalysts capable of ambient nitrogen fixation.  Active spaces ranging from [54, 54] to [65, 57] have been predicted for a quantitative description of FeMoco's electronic structure.  However, a wavefunction approach for a singlet state using a [54,54] active space would require 10$^{29}$ variables. In this work, we systematically explore the active-space size necessary to qualitatively capture strong correlation in FeMoco and two related moieties, \moietyone\ and \moietytwo. Using CASSCF and 2-RDM methods, we consider active-space sizes up to \fourteen\ and \thirty, respectively, with STO-3G, 3-21G, and DZP basis sets and use fractional natural-orbital occupation numbers to assess the level of multireference electron correlation, an examination of which reveals a competition between single-reference and multi-reference solutions to the electronic Schr\"{o}dinger equation for smaller active spaces and a consistent multi-reference solution for larger active spaces.
\end{abstract}


\section{Introduction}\label{introduction}
 \begin{figure}
 \includegraphics[scale=0.75]{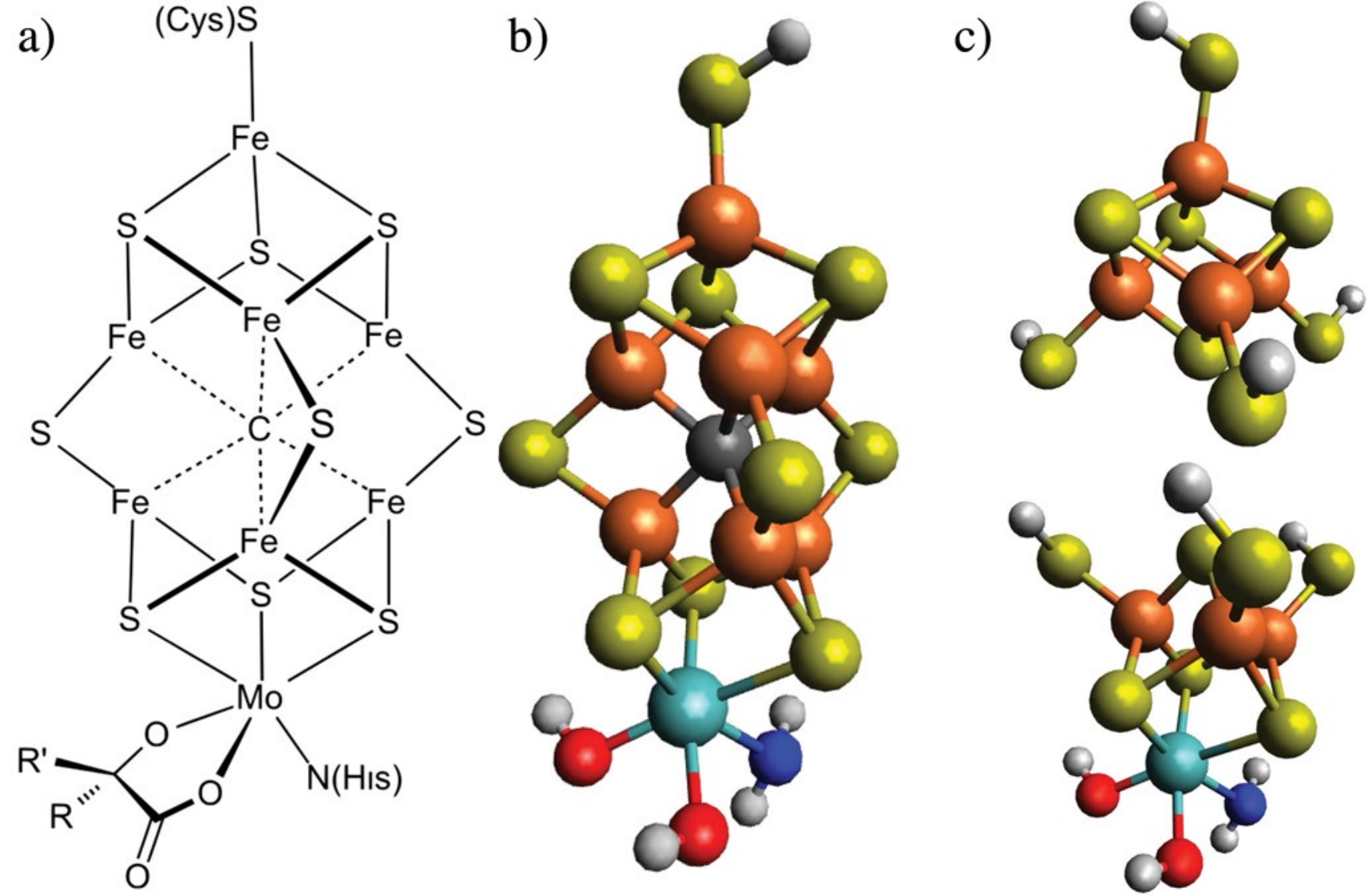} %
 \caption{\label{femoco-structure} Structures for a) native FeMoco, \FeMoco,  b) modified FeMoco considered in this work,  and c) moieties  \moietyone\ (bottom) and  \moietytwo\ (top). }
 \end{figure}

Nitrogen fixation is the energy-intensive process of converting inert atmospheric nitrogen to more reactive, bioavailable forms, such as ammonia, nitrates, etc.  Industrial nitrogen fixation is achieved primarily through the Haber-Bosch process, the metal catalyzed reduction of nitrogen to ammonia at high pressures and temperatures.   In contrast, nitrogen-fixing bacteria in the soil have the ability to convert nitrogen to ammonia under ambient conditions according to the equation\cite{cr9500545}
\begin{equation} \mathrm{N}_2 + 8\mathrm{H}^+ + 8 \mathrm{e}^- + 16 \mathrm{ATP} \longrightarrow 2\mathrm{NH}_3 + \mathrm{H}_2 + 16 \mathrm{ADP} + 16 \mathrm{P}_i. \label{reaction1}\end{equation}
Nitrogenase, the enzyme responsible for catalyzing Eq.~(\ref{reaction1}), is a two-component complex consisting of the MoFe protein, responsible for nitrogen reduction, and the Fe protein, which donates the electrons to the MoFe protein.  The redox-active site of the MoFe protein is the FeMo cofactor, or FeMoco (\FeMoco, Figure~\ref{femoco-structure}a), which has been studied extensively for decades using X-ray absorption/emission\cite{kowalska2015,lancaster2011}, electron paramagnetic resonance\cite{Rawlings25021978,pierik1993} and   M\"ossbauer\cite{Rawlings25021978,acs.inorgchem.6b02540}, and electron nuclear double resonance\cite{hoffman1982,lee1997}  spectroscopies and X-ray crystallography~\cite{Kim1677,Einsle1696}.   A mechanism for nitrogen fixation, including oxidation and protonation states of the MoFe protein, has been proposed based on available kinetic data~\cite{cr950055x,cr400641x}.

Density functional theory (DFT) has been used to calculate nitrogen binding and propose mechanism intermediates~\cite{Solr-8855734}.  Broken symmetry (BS) DFT~\cite{Noodleman1992} has been used to help elucidate X-ray  and M\"ossbauer spectroscopic data to assign oxidation states of FeMoco's Fe and Mo atoms, assignments that are not yet fully settled~\cite{C4SC00337C,acs.inorgchem.6b02540}.   It is also known that in the absence of sufficient Mo soil concentrations, homologous V and Fe based forms of the nitrogenase cofactor, FeVco and FeFeco, respectively, have evolved, with  FeMoco $>$ FeVco $>$ FeFeco in terms of efficiency of nitrogen fixation~\cite{cr950057h,Eady2003}.  FeVco, however, has been shown to be more efficient at the reduction of carbon monoxide to small hydrocarbons~\cite{Lee642}.

So while much work has been done to understand FeMoco's electronic structure, open questions remain. DFT is capable of describing dynamic correlation, but a thorough description of FeMoco's redox properties should include an understanding of static, or multireference, correlation effects as well.  Multireference correlation, which arises when more than one determinant must be included in the reference wavefunction to account for contributions from (nearly) degenerate electron configurations,  can have considerable consequences for a molecule's redox properties. For example, in organometallic redox reactions, it is possible that the organic ligand, not the metal, plays the role of the electron donor or acceptor in a process known as {\em ligand non-innocence}~\cite{JORGENSEN1966164,cs200660v,Ray2007}.   Recent studies involving ligand non-innocence in manganese porphyrins and vanadium oxo complexes have shown that methods that fail to capture multireference electron correlation can mistakenly attribute a redox event to being metal-centered~\cite{C6CP07563K,acs.jpclett.5b02547, acs.jpca.7b09567}.  In separate studies on polyaromatic hydrocarbons, multireference correlation was found to give rise to increasing polyradical character with increasing size for both linear chains and two-dimensional geometries\cite{jp2017192}.

In this work, we study strong electron correlation in FeMoco using two multireference methods: the configuration interaction complete-active-space self-consistent-field (CI-CASSCF) method~\cite{siegbahn1981,AvdChemPhys69} and the variational two-electron reduced density matrix (V2RDM) method~\cite{cr2000493,RDM-Mechanics,mazziotti042113,nakata2001,mazziotti062511,zhao2095,mazziotti213001,mazziotti10957,cances064101,mazziotti032501,mazziotti249,gidofalvi2008,shenvi2010,greenman164110,verstichel213001,mazziotti263002,fosso2260,mazziotti153001}. Strictly speaking, both are complete-active-space methods in which those electrons and orbitals thought to exhibit the greatest entanglement are treated as explicitly correlated. Orbital rotations are performed so as to minimize the energy in a self-consistent fashion.  In the CI-CASSCF approach, the $N$-electron wavefunction is expressed as a linear combination of determinants constructed from active orbitals. Because expansion coefficients and electronic energies are determined via matrix diagonalization, CI-CASSCF scales exponentially with the number of active orbitals and is limited to smaller active-space sizes.  In the V2RDM approach, however, the energy is expressed as a functional of the two-electron reduced density matrix, which scales only polynomially with the number of active orbitals, allowing larger active spaces to be considered. To decrease the computational cost, we consider a modified FeMoco in which those atoms and moieties that anchor the cofactor into the protein are replaced with H (Figure~\ref{femoco-structure}b).   We also consider two smaller cluster moieties, \moietyone\ and \moietytwo\ (Figure~\ref{femoco-structure}c).  In each case, the aim of the work is to study systematically the emergence of static correlation using increasingly larger active spaces and basis sets. To our knowledge, this work represents the largest active-space calculation of FeMoco and is a promising step forward in understanding the role of strong static correlation in its electronic structure.

\section{Theory}\label{theory}

In this section, we describe the CI-CASSCF and V2RDM calculations.  Both methods are complete-active-space methods defined by \AS, where $N_a$ is the number of active electrons and $r_a$ is the number of active orbitals.

CI-CASSCF calculations employed in this work proceed by (i) selecting an underlying basis-set representation and an initial active-space size \AS, (ii) expressing the many-electron wavefunction as a linear combination of determinants constructed from the active orbitals followed by matrix diagonalization to obtain expansion coefficients and energies, (iii) performing an orbital rotation between active and inactive spaces to generate an improved set of active orbtials, (iv) repeating steps (ii) and (iii) in a self-consistent fashion until convergence, and (v) performing a Mulliken population analysis.  Due to the complexity FeMoco, choosing which orbitals to include in the active space is a challenge.  For the systematic study of basis and active-space size, we choose to use the Hartree-Fock orbitals, the first $(N - N_a)/2$ orbitals designated as core and the next $r_a$ designated as active orbitals,  where $N$ is the total number of electrons in the system.  The matrix diagonalization in step (ii) is computationally expensive and scales exponentially with system size as $O(r_a^{N_a})$, limiting its use to smaller active-space sizes. We used the PYSCF package\cite{pyscf} to carry out the CI-CASSCF calculations and Mulliken population analysis and GMolden\cite{gmolden} to visualize the natural orbtials.

The V2RDM calculations proceed in a similar fashion, except that the expensive matrix diagonalization in step (ii)  is replaced by a minimization of the ground-state energy, $E$, as a functional of the two-electron reduced density matrix, \rdm:
\begin{equation}
     E = \mathrm{Tr}[^2K\ ^2D]
\end{equation}
where $^2K$ is the 2-electron reducted Hamiltonian matrix.  The minimization is achieved using semidefinite programming~\cite{mazziotti213001,mazziotti10957,mazziotti249,nakata2001, mazziotti2011}, a constrained optimization approach based on the $p$-particle, $N$-representability conditions~\cite{shenvi2010} that ensure that the 2-RDM corresponds to an $N$-electron wavefunction.  In this work, we use 2-positivity conditions, which require the 2-RDM, as well as the 2-hole RDM, $^2Q$, and the particle-hole RDM, $^2G$, to be positive semidefinite:
\begin{eqnarray}
^2D \succeq 0 \\
^2Q \succeq 0 \\
^2G \succeq 0,
\end{eqnarray}
whose matrix elements are given by
\begin{eqnarray}
^2D_{k,l}^{i,j} = \bra \psi \ahatd{i}\ahatd{j}\ahat{l}\ahat{k}\ket\psi \\
^2Q_{k,l}^{i,j}  = \bra \psi \ahat{i}\ahat{j}\ahatd{l}\ahatd{k}\ket\psi \\
^2G_{k,l}^{i,j} = \bra \psi \ahatd{i}\ahat{j}\ahatd{l}\ahat{k}\ket\psi.
\end{eqnarray}
where the creation and annihilation operators correspond only to the active-space orbitals.  The use of $D$, $Q$, and $G$ conditions gives rise to $O(r_a^6)$ scaling, as opposed to the exponential scaling of CASSCF allowing for larger active-space sizes to be considered.  The use of $p < N$ positivity conditions gives rise to a {\em lower} bound for the exact ground-state energy. Higher order $p$-positivity conditions can be imposed to improve upon this lower bound at the cost of increased computational effort.

Predicting the active-space size necessary to describe accurately the system's multireference correlation remains a challenge.  In some cases, chemical intuition can be used, but for systems like FeMoco, a large asymmetric complex of multiple transition metals and sulfur atoms, chemical intuition can be completely lacking.  Estimations of \AS\ based on the number of valence electrons and orbitals can overestimate the size of the active space necessary, as atoms are intrinsically more open shell than molecules. Reiher {\em et al.}  considered quantum algorithms and resources necessary to elucidate nitrogenase's reaction mechanism, predicting active spaces ranging from  [54, 54] to [65, 57], depending on the spin and charge state considered~\cite{Reiher18072017}.   A CAS approach for a singlet state using a [54,54] active space would correspond to 10$^{29}$ configuration state functions!  In this work, we explore the active-space size necessary to capture qualitatively the dominant correlation by systematically increasing the active space and using fractional spatial natural-orbital occupation numbers to assess the level of multireference electron correlation.  Natural-orbital occupation numbers for spatial orbitals should range from 0 to 2, and while the cutoff criterion for correlated versus uncorrelated electrons is somewhat arbitrary, we choose the range 0.2 to 1.8 to indicate strong electron correlation.  Natural orbitals with occupation numbers less than 0.2 or greater than 1.8 are considered to have more virtual-orbital or core-orbital character, respectively. As an additional means to assess the level of correlation, we  calculate the von Neumann entropy\cite{vonNeumann55}, $S_1$, defined as
\begin{equation}
S_1 = -\mathrm{Tr} (^1D\mathrm{ln} (^1D)) = - \sum_i^{r_a} \frac{n_i}{2}\mathrm{ln}\left(\frac{n_i}{2}\right),\label{von-Neumann}
\end{equation}
where $n_i$ is the occupation number of the $i$th active orbital.  The von Neumann entropy is zero in the absence of electron correlation and increases with more fractional occupation numbers between 0 and 2.

To this end, we consider active-space sizes \six, \ten, \fourteen, \eighteen, \twentytwo, \twentysix, and \thirty.  The exponential scaling of the CI-CASSCF method limits its use to active spaces up to \fourteen\ while the V2RDM approach, with its $r_a^6$ scaling, allowed for active spaces up to \thirty.  Consider, the \thirty\ active space would require less than $10^6$ variables in the 2-RDM to represent the corresponding singlet wavefunction with its $10^{15}$ variables.   The underlying basis representation can also influence the ability of a given active space to capture electron correlation. We explore increasing basis-set sizes, including the minimal STO-3G basis and two double-zeta basis sets: 3-21G\cite{321G,321G2,321G3,321G4} and DZP\cite{DZP, DZP2,DZP3}.   Coordinates for FeMoco were based on those used in Ref.~\citenum{Reiher18072017} with H-atoms substituted for anchoring R-groups. (See Supporting Information, Table S1.)   The structure considered differs from native FeMoco only in those atoms that would anchor the subunit into the parent protein and therefore provides a useful probe into the extent of electron correlation in FeMoco.  For added simplicity, all calculations correspond to neutral compounds in the singlet S=0 state, as opposed to the native charged, S=3/2 state~\cite{Munck1975, Zimmermann1978}.  As an additional probe into the electronic structure of FeMoco, we also consider the top and bottom halves of Femoco separately by replacing the central carbon with a sulfur and replacing sulfur-iron bonds with sulfur-hydrogen bonds, as seen in  Figure~\ref{femoco-structure}b.  We refer to the bottom and top halves as \moietyone\ and \moietytwo, respectively. (See Supporting Information, Tables S2 and S3.)

\section{Results}\label{results}
In this section, we present results based on V2RDM calculations for FeMoco (Figure~\ref{femoco-structure}b) as well as for moieties \moietyone\ and \moietytwo\ (Figure~\ref{femoco-structure}c).   In order to study the effects of active-space size and basis set on the multireference electron correlation seen in the system, we consider a set of increasingly large active space-sizes, including \AS\ = \six, \ten, \fourteen, \eighteen, \twentytwo, \twentysix, and \thirty\ in the STO-3G, 3-21G, and DZP basis sets.  We also present results for CI-CASSCF calculations for \six, \ten, and \fourteen\ in each basis set. Larger active spaces were intractable.

     \begin{figure}
   	\includegraphics[scale=1.0]{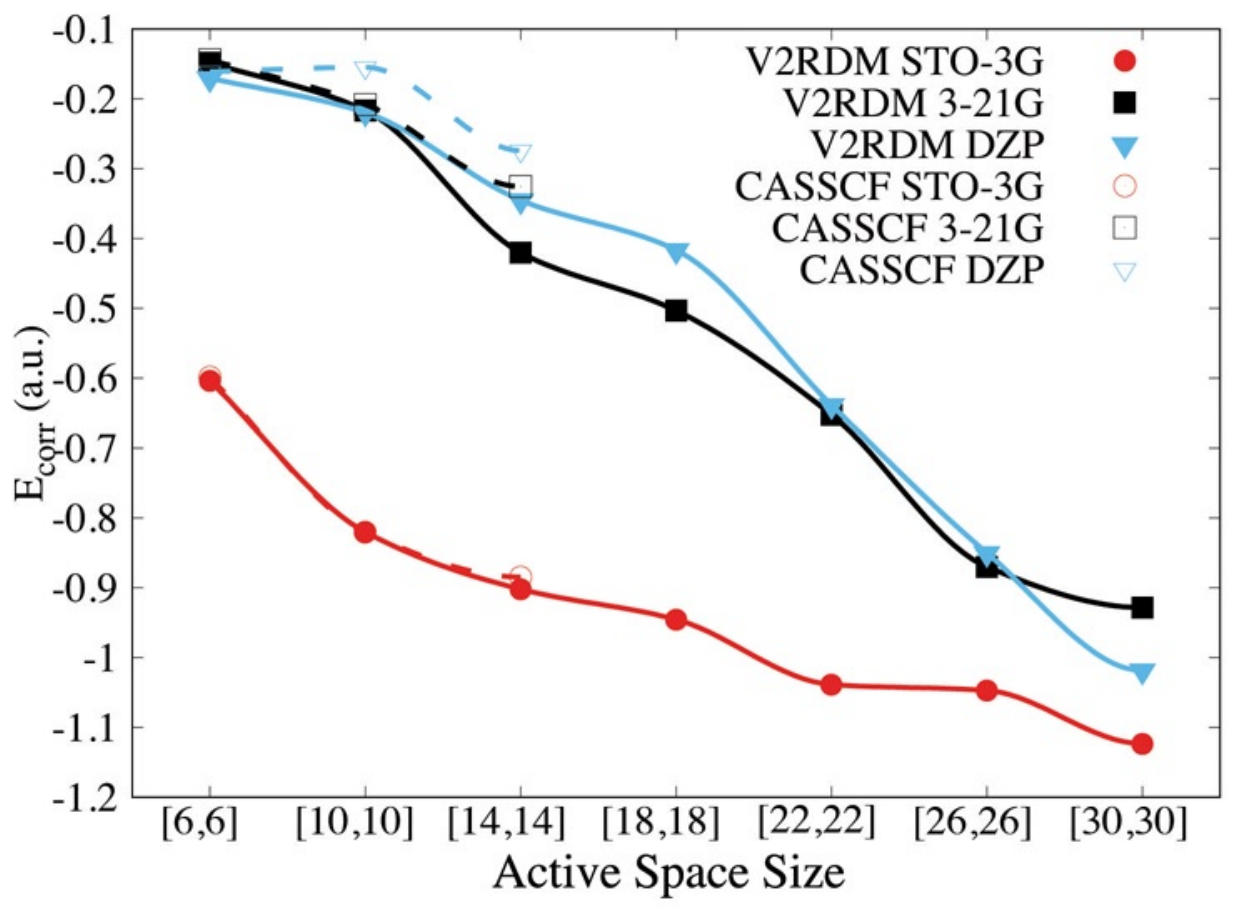} %
 	\caption{\label{correlation-energies} FeMoco correlation energies (a.u.) for STO-3G (red, circles), 3-21G (black, squares), and DZP (blue, triangles) basis sets and increasing active-space size. V2RDM energies correspond to solid lines and filled symbols while CASSCF correspond to dashed lines and analogous open symbols. The HF energies are -16850.9947 a.u., 	-16948.2674 a.u., and	-17030.7393 a.u. for the STO-3G, 3-21G, and DZP basis sets, respectively.  All calculations correspond to an initial active space composed of restricted HF orbitals.}
      \end{figure}
\begin{figure}
 \includegraphics[scale=0.8]{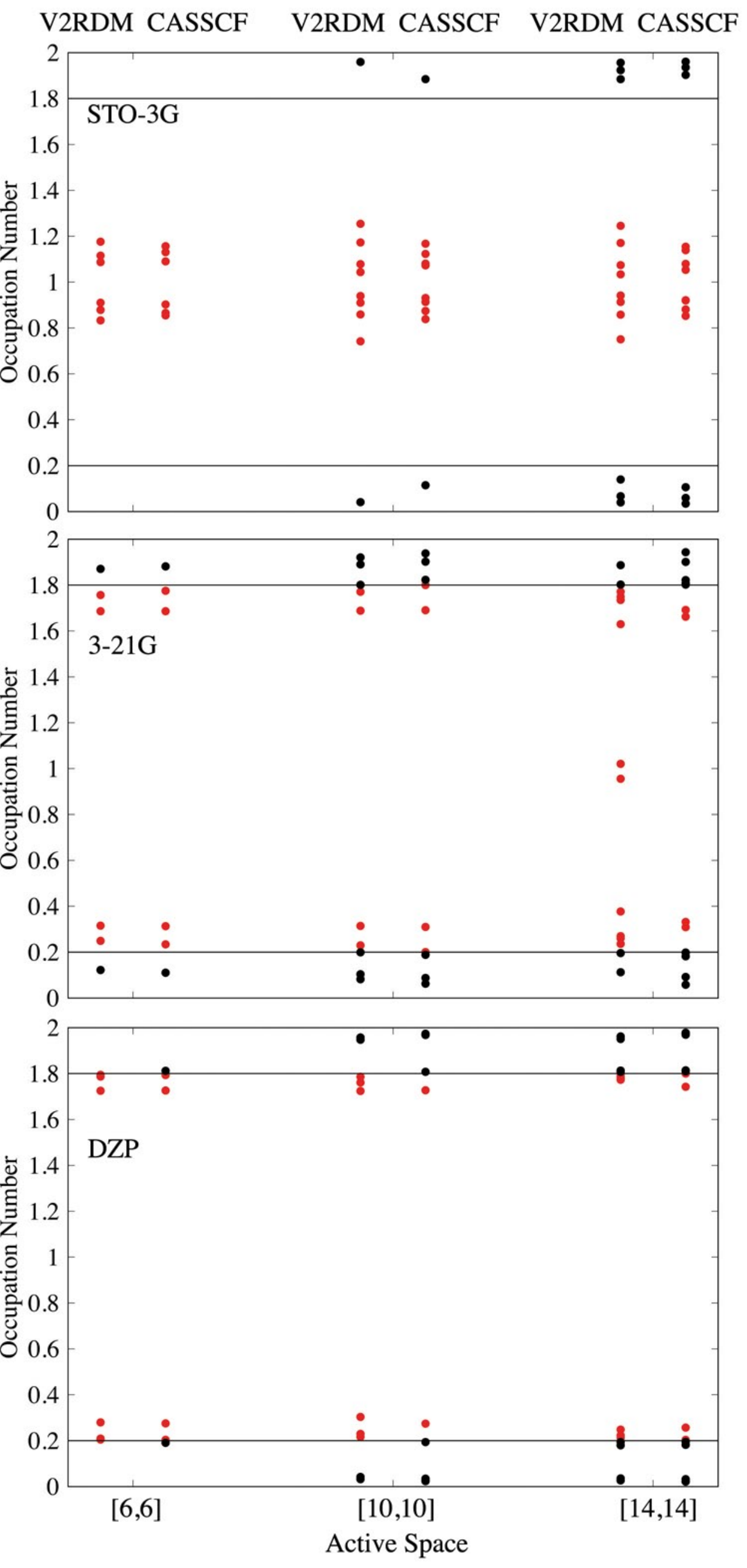} %
 \caption{\label{casscf-ONs} Occupation numbers for V2RDM and CASSCF in STO, 3-21G, and DZP basis sets for \six, \ten, and \fourteen.}
 \end{figure}

We show in Figure~\ref{correlation-energies} correlation energies (in atomic units) for FeMoco calculated with CASSCF (up to \fourteen) and V2RDM (up to \thirty) for the three basis sets, where the correlation energy is defined as the difference between the CASSCF or V2RDM energy and the Hartree Fock (HF) energy in the same basis set. (Correlation energies also provided in Tables S4 and S5 in the Supporting Information.)  The HF energies for the STO-3G basis set ($E_\mathrm{HF} = -16850.9947$ a.u.), the 3-21G basis set ($E_\mathrm{HF} = -16948.2674$ a.u.), and the DZP basis set ($E_\mathrm{HF} = -17030.7393$ a.u.)  are consistent with the notion that energies should decrease for improved basis-set calculations.  We note that correlation energies from V2RDM and CASSCF calculations in the STO-3G basis set differ very little but begin to deviate for the double-zeta 3-21G and DZP basis sets.  This difference can be attributed to the use of 2-positivity conditions.   Higher-order positivity conditions would rectify the difference in correlation energies with a computational cost. However, it is important to note that the 2-positivity conditions are sufficient to capture the key multireference correlation in the system, as evidenced by the similarity in the fractional occupations numbers in Figure~\ref{casscf-ONs}.  Correlation energies for moieties \moietyone\ and \moietytwo\ show similar trends to FeMoco.

The extent of electron correlation was determined by looking at the advent of fractional occupation numbers of the active natural orbitals.    Figure~\ref{casscf-ONs} contains a plot of FeMoco occupation numbers for the smaller \six, \ten, and \fourteen\ active spaces for both the V2RDM and CASSCF methods in the three basis sets.  For each active-space size, the first column of data points corresponds to the V2RDM results and the second to the CASSCF results. Occupation numbers between 0.2 and 1.8 are considered to correspond to strongly correlated electrons and are plotted in red while occupation numbers outside this range are plotted in black.
While the two methods render a consistent description, the minimal STO-3G and double-zeta basis sets give rise to conflicting multireference and single-reference descriptions, respectively, for these smaller active spaces.

\begin{figure}
 \includegraphics[scale=0.8]{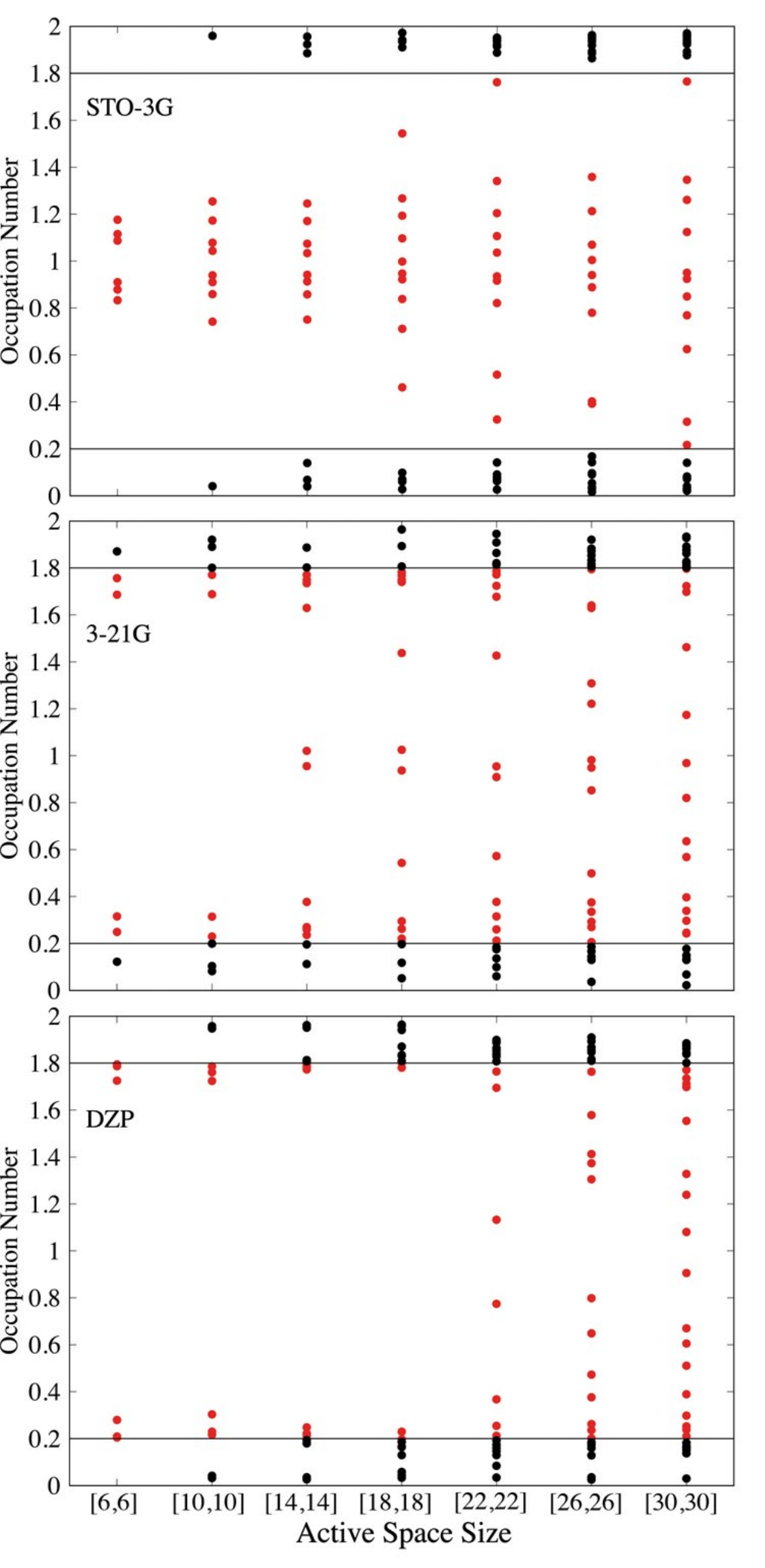} %
 \caption{\label{femoco-ONs} Occupation numbers for FeMoco using V2RDM for increasing active-space size for STO-3G, 3-21G, and DZP basis sets. The $x$-axis is the same for each plot and is displayed at the bottom.}
 \end{figure}

With the V2RDM method, we were able to consider active spaces up to \thirty.  Figure~\ref{femoco-ONs} collates occupation numbers for a larger range of active spaces up to \thirty\ for the STO-3G, 3-21G, and DZP basis sets.  Table~\ref{table-vonNeumann} contains corresponding von Neumann entropies calculated using Eq.~(\ref{von-Neumann}).  It is clear that the active-space composition as well as the underlying basis-set representation can affect the extent of correlation captured in the system.
\begin{table}[h]
\caption{\label{table-vonNeumann}  Calculated von Neumann entropies for FeMoco.}
\smallskip
\smallskip
\begin{tabular}{ccccc}
&	STO-3G	&	3-21G	&	DZP 	\\
\hline
\six\	&	2.05	&	1.04	&	1.07	\\
\ten\	&	2.82	&	1.49	&	1.31	\\
\fourteen\	&	3.22	&	2.85	&	1.85	\\
\eighteen\	&	3.80	&	3.41	&	2.12	\\
\twentytwo\	&	4.03	&	4.08	&	3.51	\\
\twentysix\	&	4.26	&	5.18	&	4.70	\\
\thirty\	&	4.55	&	5.53	&	5.95	\\

\end{tabular}
\end{table}
The minimal STO-3G basis gives rise to a consistent multireference solution for all active-space sizes. The double-zeta 3-21G basis set provides enough flexibility to stabilize effectively the HF orbitals for smaller active-space sizes but gives rise to a more multireference solution for the \fourteen\ and larger active space.  This is even more pronounced when polarization basis functions are introduced in the DZP basis set, which maintains a single-reference character until \twentytwo.  For the STO-3G and 3-21G basis sets,  the number of correlated electrons at the largest active-space size becomes fairly constant and is fewer than half of the active electrons.  While we cannot say that we have reached a similar level of convergence for the DZP basis set, the 3-21G results suggest that we may be close to a qualitatively converged description of multireference correlation in the system. The lack of fractional occupation numbers for smaller active-space sizes for DZP also suggests that correlated methods, such as CASSCF, with insufficient active spaces can fail to represent the electron correlation present in the system with possible consequences in predicting redox behavior.  The ability of the V2RDM method to treat larger active spaces self-consistently provides a powerful tool in describing electron correlation in larger organometallic complexes.

\begin{figure}
 \includegraphics[scale=1]{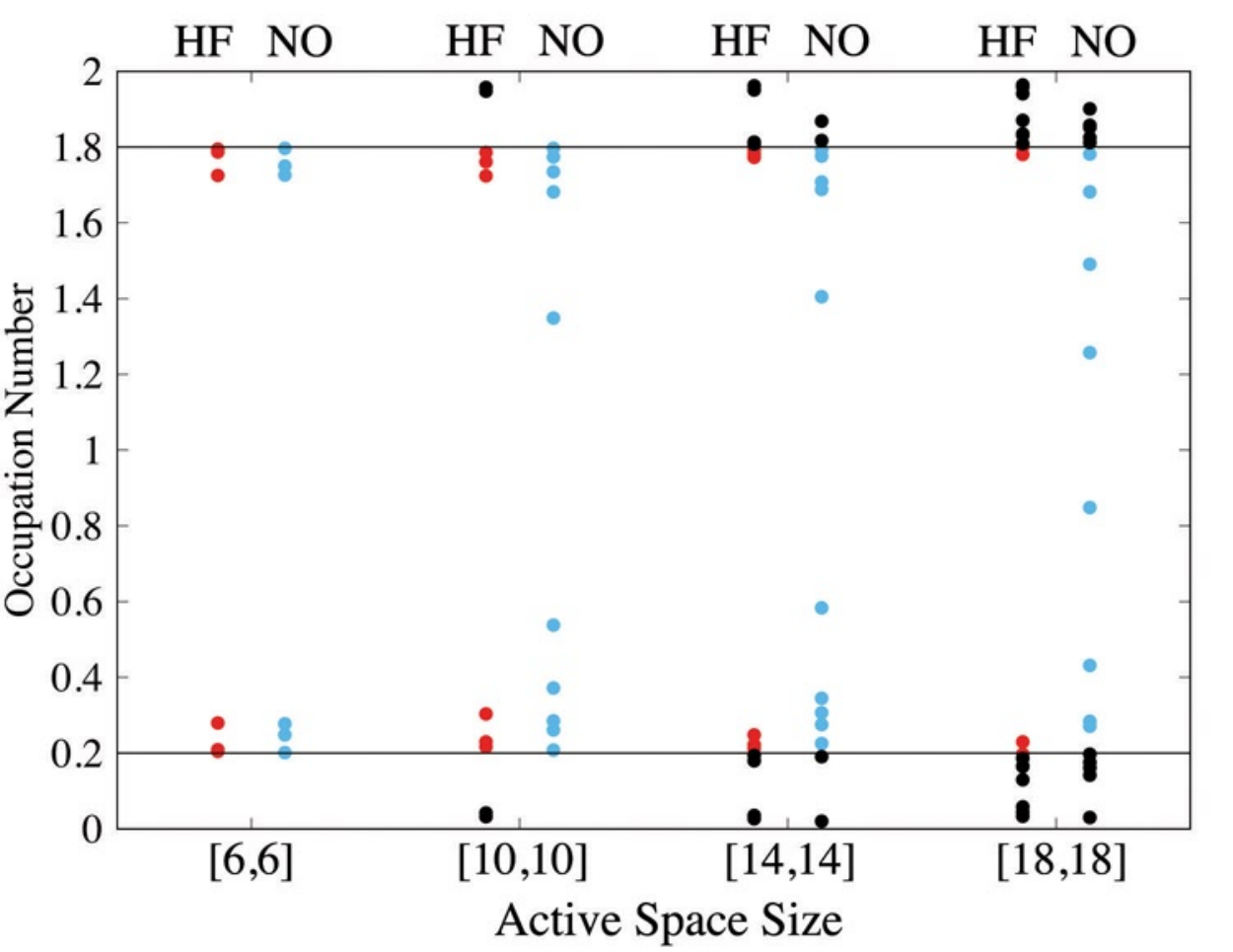} %
 \caption{\label{seeded-ONs} Occupation numbers for \six, \ten, \fourteen, and \eighteen\ active spaces seeded with HF orbitals (red) and natural orbitals (blue) from the \twentytwo\ active space calculation.}
 \end{figure}

In order to determine if calculations were converged with respect to basis, we ran \six, \ten, and \fourteen\ active-space calculations with the TZP basis~\cite{TZP,TZP2,TZP3} and found orbital occupation numbers to be converged with single-reference solutions persisting.  In order to investigate the effect of the initial active space orbitals, we first compared [6, 10], [8, 10], [10, 10], [14, 10] and [18,10] active spaces in the DZP basis.  Active spaces with $N_a < r_a$ are composed of more virtual HF orbitals than occupied HF orbitals, and vice verse.  In each case, single-reference character persisted with the [10, 10] active space giving rise to the lowest energy.  We also considered \six, \ten, \fourteen, and \eighteen\ active spaces composed of correlated natural orbitals (NOs) from the \twentytwo\ active-space calculation.   In each case, the CAS-CI energy was greater with more single-reference character when using the NOs from the \twentytwo\ active space.  However, subsequent orbital rotations in the SCF procedure lowered energies and gave rise to more multireference character than those calculations seeded with HF orbitals for all but the \six\ active space. Correlation energies are provided in Table S5 in the Supporting Information. Figure~\ref{seeded-ONs} compares the occupation numbers calculated using HF orbitals (red) and natural orbitals (blue) from the \twentytwo\ active-space calculation.  Recalculated von Neumann entropies for \six, \ten, \fourteen, and \eighteen\ are 1.10, 2.18, 2.62, and 3.26, respectively.
%
%
%
These results indicate that for smaller active spaces, multiple local minima exist with varying degrees of multireference character, and the initial active space choice can affect in which minimum the solution converges.  These results also indicate that orbital rotations are absolutely necessary regardless of which orbitals constitute the initial active space. However, even with orbital rotations, smaller active-space calculations  initially seeded with correlated natural orbitals from larger active-space calculations do not give the same description of correlation as the larger active-space calculation.  In the case of smaller active spaces, the use of natural orbitals converged from perhaps larger, coarser calculations can help reveal the multireference inherent in the system.

\begin{figure}
 \includegraphics[scale=1]{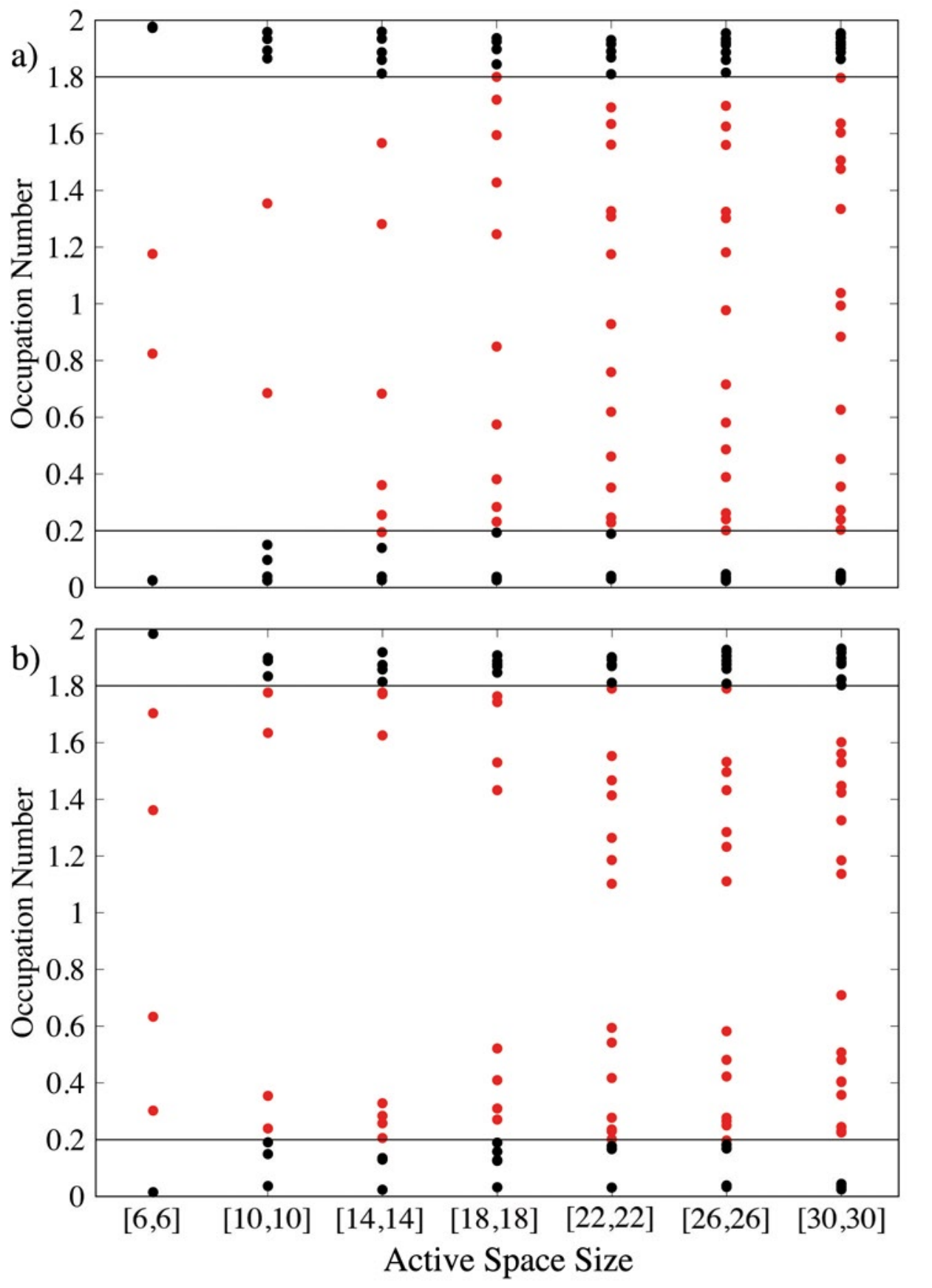} %
 \caption{\label{cubane-ONs} Occupation numbers for a) \moietyone\ moiety and b) \moietytwo\ moiety in DZP basis}
 \end{figure}

Figures~\ref{cubane-ONs}a and \ref{cubane-ONs}b  depict a similar treatment of natural-orbital occupation numbers for moieties \moietyone\ and \moietytwo, respectively, in the DZP basis as a function of active-space size.
For the smaller \moietyone\ and \moietytwo\ moieties, we see a similar competition between single-reference and multireference solutions for the molybdenum containing moiety, \moietyone, and a rough convergence of the fractional occupation numbers for both moieties by the [22,22] active space in the DZP basis. We also see more electron correlation for the individual moieties than for the whole, even for the smaller active spaces. Bond formation therefore can give rise to fewer overall correlated electrons.

\begin{figure}
 \begin{center}
 \includegraphics[scale=0.8]{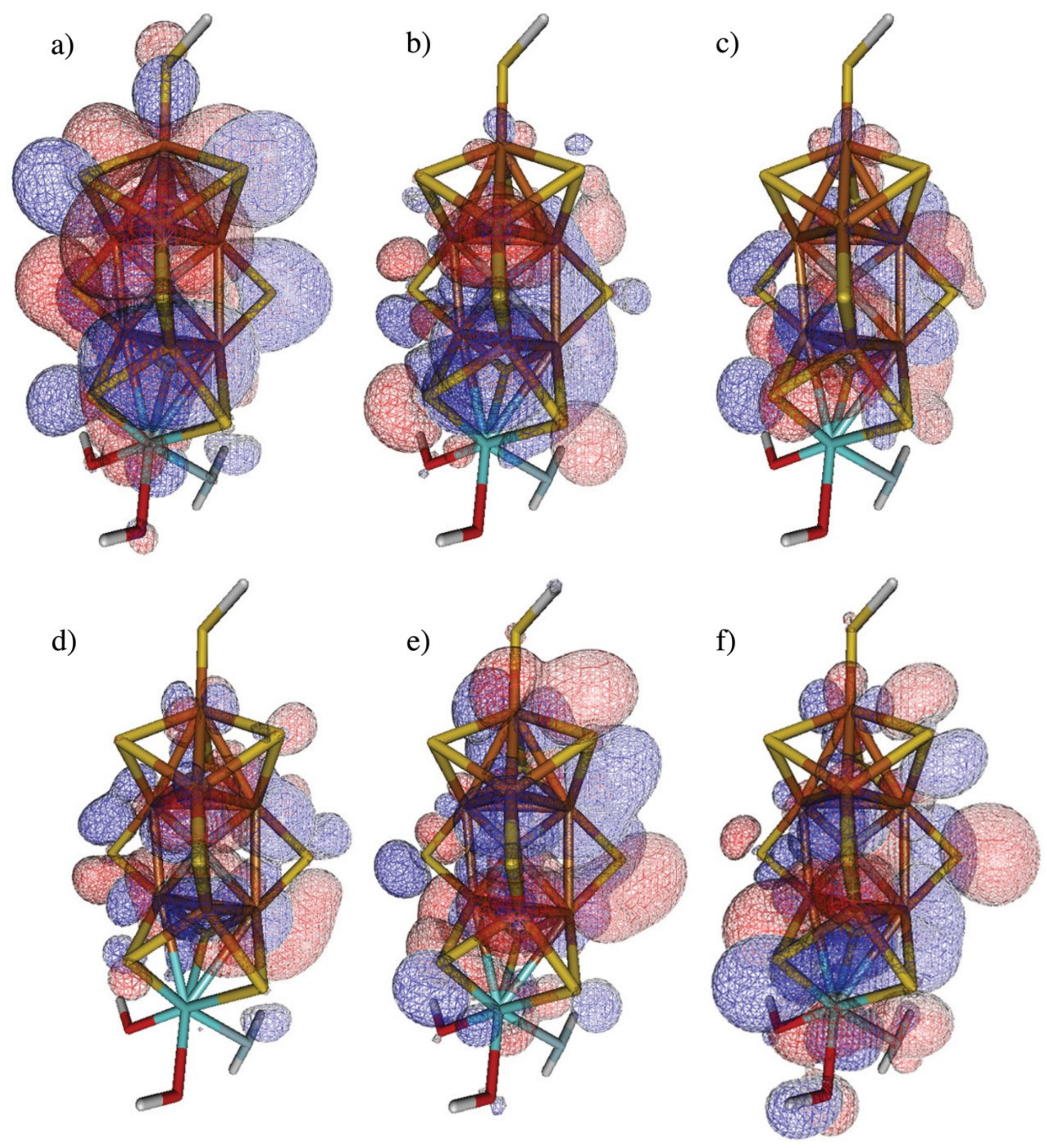} %
 \caption{\label{NOs-DZP} Natural orbitals for FeMoco corresponding to occupation numbers a) 1.77, b) 1.55, c) 1.33, d) 0.91, e) 0.67, and f) 0.24 in the DZP basis.}
\end{center}
 \end{figure}

\begin{figure}
 \begin{center}
 \includegraphics[scale=0.525]{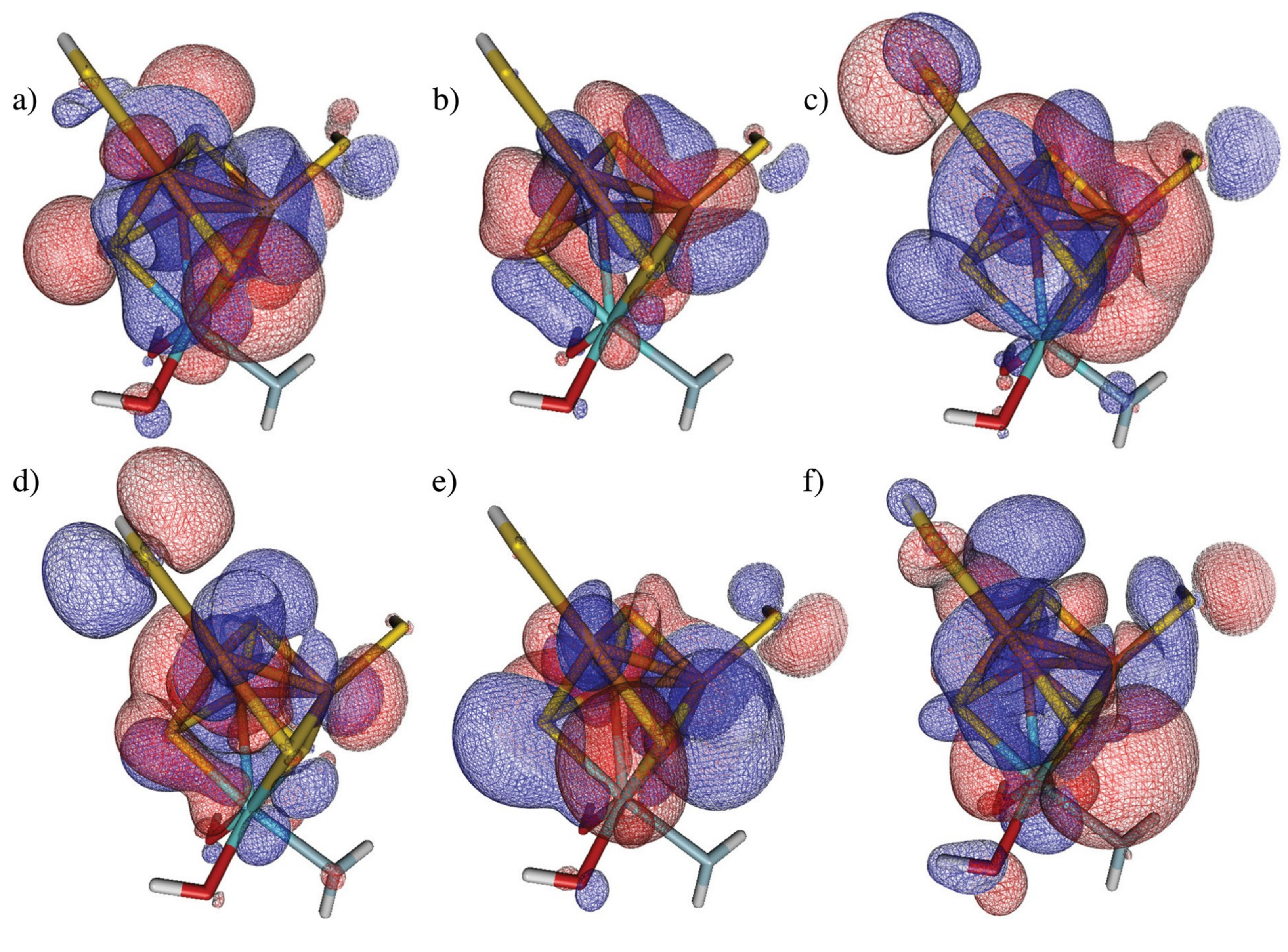} %
 \caption{\label{cubane1-NOs} Natural orbitals for moiety I corresponding to occupation numbers a) 1.60, b) 1.48, c) 1.33, d) 0.99, e) 0.63, and f) 0.36 in the DZP basis.}
 \end{center}
 \end{figure}

As an additional gauge of the effects of multireference electron correlation on FeMoco's electronic structure, we show in Table~\ref{mulliken} Mulliken charges for all three structures calculated with HF and with V2RDM in a \thirty\ active space, both in the DZP basis. The table is arranged such that analogous atoms in the moieties are in the same row as in the FeMoco structure for easy comparison.  Note that the C atom in FeMoco is replaced with an S-H  in both moieties to maintain a singlet configuration.
\begin{table}[h]
\caption{\label{mulliken}  Mulliken charges for Fe, Mo, and S in the DZP basis and [30,30] active space.}
\smallskip
\smallskip
\begin{tabular}{c|ccc|ccc}

Atom & FeMoco &  \moietyone & \moietytwo & FeMoco &  \moietyone & \moietytwo  \\
	 & (HF) & (HF) & (HF) & (V2RDM) & (V2RDM) & (V2RDM) \\
\hline
Fe1	&	1.08942	&		&	1.10943	&	0.46494	&		&	0.38516	\\
S1	&	-0.73244	&		&	-0.78483	&	-0.34418	&		&	-0.19178	\\
S2	&	-0.778	&		&	-0.60084	&	-0.36613	&		&	-0.16883	\\
S3	&	-0.73931	&		&	-0.76215	&	-0.30013	&		&	-0.11723	\\
Fe2	&	1.0659	&		&	0.98745	&	0.64095	&		&	0.40977	\\
Fe3	&	1.19877	&		&	1.11236	&	0.6791	&		&	0.37605	\\
Fe4	&	1.12446	&		&	0.96818	&	0.65445	&		&	0.39176	\\
C/S	&	-1.76377	&	-0.27864	&	-0.26555	&	-1.33811	&	0.08011	&	0.06908	\\
S4	&	-0.64224	&	-0.56727	&	-0.65468	&	-0.35297	&	-0.4246	&	-0.40623	\\
S5	&	-0.68626	&	-0.6775	&	-0.4758	&	-0.25305	&	-0.4361	&	-0.41797	\\
S6	&	-0.67731	&	-0.86157	&	-0.65245	&	-0.30585	&	-0.38741	&	-0.3955	\\
Fe5	&	1.17227	&	1.03452	&		&	0.52043	&	0.33275	&		\\
Fe6	&	0.99114	&	1.03212	&		&	0.46296	&	0.40378	&		\\
Fe7	&	0.88262	&	0.96051	&		&	0.49618	&	0.2496	&		\\
S7	&	-0.68347	&	-0.72916	&		&	-0.23225	&	-0.09239	&		\\
S8	&	-0.58731	&	-0.46001	&		&	-0.36287	&	-0.05323	&		\\
S9	&	-0.61297	&	-0.75982	&		&	-0.26157	&	-0.29959	&		\\
Mo	&	1.45167	&	2.18944	&		&	1.17767	&	1.11051	&		\\
\end{tabular}
\end{table}
In each case, we see the expected positively charged transition metals, with Mo in a higher oxidative state compared to the Fe atoms, and negatively charged sulfur atoms. But we also notice that multireference correlation manifests itself in more electron density on the transition metals, resulting in more reduced Mo and Fe atoms and more oxidized S atoms than in the single-reference HF picture.  In FeMoco, this is more pronounced for Fe, and in \moietyone, this is more pronounced for the Mo.   A similar story unfolds in looking at plots of correlated natural orbitals for FeMoco and the \moietyone\ moiety, plotted in Figures~\ref{NOs-DZP}a-\ref{NOs-DZP}f and Figures~\ref{cubane1-NOs}a-\ref{cubane1-NOs}f, respectively, each containing a representative selection of correlated natural orbitals.
Not surprisingly, correlated orbitals in both entities show considerable amplitude on the Fe, Mo, and S atoms, but the \moietyone\ moiety shows more natural-orbital amplitude on the Mo throughout the range of fractional occupation numbers, giving rise to a more reduced molybdenum in the \moietyone\ moiety.  Natural orbitals for the \moietytwo\ moiety, not presented, show considerable amplitude on the sulfurs and irons throughout the range of fractional occupation numbers.

\section{Discussion and Conclusions}\label{discussion}
The cofactor FeMoco, \femoco, has been shown to be the active catalytic site responsible for the reduction of molecular nitrogen (N$_2$) to ammonia in the naturally occurring nitrogenase protein.  A detailed understanding of its electronic structure, including strong electron correlation, is key to designing mimic catalysts capable of ambient nitrogen fixation.   This work pursues FeMoco's electronic structure from a multireference perspective using the CI-CASSCF and V2RDM methods. We systematically explore the effect of active-space size and basis-set representation on the extent of strong static correlation  in FeMoco and two moieties \moietyone\ and \moietytwo\ and find that both active-space size and basis-set representation can give rise to a competition between single-reference and multireference solutions when using HF orbitals in the initial active space.  Double-zeta basis sets 3-21G and DZP favor single-reference solutions for smaller active spaces but gave rise to correlated, multireference solutions at larger active space sizes.  This competition can be mitigated somewhat by using correlated natural orbitals from larger active space calculations. At \thirty\ in the DZP basis, calculated Mulliken charges and plots of natural-orbitals with fractional occupation numbers reveal static correlation gives rise to more reduced molybdenum and iron atoms compared to the single-reference HF solution, a consequence that should affect FeMoco's redox properties.

Recent studies involving ligand non-innocence have shown that insufficient active space sizes can fail to capture key multireference correlation and lead to qualitatively different descriptions of redox events.  Calculations of the reduction of vanadium(IV) oxo complex using the V2RDM method and an active space of [12, 10] supported a metal centered electron addition giving rise to an elusive low-valent vanadium (III) oxo complex while a larger [42, 40] active-space calculation revealed that electron entanglement leads to a ligand-centered addition.  A very similar result was seen in a study involving manganese superoxide dismutase mimics for which a CI-CASSCF calculation using an [14, 14] gives rise to a metal-centered electron addition while a V2RDM calculation with a larger [30, 30] active space points to ligand-centered addition and ligand non-innocence.  While the current work does not consider the reduction or oxidation of FeMoco,  competing single-reference and multireference solutions at smaller and larger active spaces, respectively, reinforce the need for a qualitatively correct description of static correlation to describe accurately redox events.

\begin{acknowledgement}

D.A.M. gratefully acknowledges the United States National
Science Foundation Grant CHE-1565638, the United States-
Army Research Office (ARO) Grants W911NF-16-C-0030 and
W911NF-16-1-0152, and the United States Air Force Office of
Scientific Research Grant FA9550-14-1-0367.

\end{acknowledgement}

\begin{suppinfo}\label{suppinfo}

We include in Supporting Information the geometries and correlation energies for all calculations involving \femoco, \moietyone, and \moietytwo .

\end{suppinfo}

\bibliography{refs}

\end{document}